\documentclass[11pt]{amsart}

\usepackage[utf8]{inputenc}
\usepackage[T1]{fontenc}
\usepackage{mlmodern}

\usepackage{amssymb}
\usepackage{mathtools}
\usepackage{amscd}
\usepackage[mathscr]{eucal}
\usepackage{enumitem}
\usepackage[hmargin=1in,vmargin=1in]{geometry}

\usepackage{graphicx}
\usepackage{xcolor}

\usepackage[roman]{sublabel}

\usepackage[labelfont=bf,font=small]{caption}
\usepackage{subcaption}

\usepackage[noadjust]{cite}
\usepackage[foot]{amsaddr}

\usepackage{placeins}
\usepackage{bm}
\usepackage{siunitx}

\usepackage{booktabs}

\usepackage{comment}

\usepackage{algorithm2e}
\RestyleAlgo{ruled}
\definecolor{CitePurple}{RGB}{128,9,158}
\definecolor{CiteBlue}{RGB}{2,95,176}
\definecolor{LinkRed}{rgb}{0.7,0,0}
\usepackage[colorlinks=true,
            citecolor=CitePurple,
            linkcolor=LinkRed,
            urlcolor=CiteBlue]{hyperref}

\theoremstyle{plain}

\theoremstyle{definition}

\newcommand{\revision}[1]{{\color{black} #1}}

\DeclareMathOperator{\E}{\mathbb{E}}
\DeclareMathOperator{\N}{\mathbb{N}}
\DeclareMathOperator{\Z}{\mathbb{Z}}
\DeclareMathOperator{\Prob}{\mathbb{P}}
\DeclareMathOperator{\cv}{cv}
\DeclareMathOperator{\ev}{\mathbb{E}}

\DeclareMathAlphabet{\mathbbold}{U}{bbold}{m}{n}

\begin{document}

\raggedbottom

\title{Distributed delay stabilizes bistable genetic networks}

\author{Sean Campbell$^{1}$}

\author{Courtney C.\ White$^{2}$}

\author{Amanda M.\ Alexander$^{*,3}$}

\thanks{$^{*}$Corresponding author:
Contact \href{mailto:amalexander.math@gmail.com}{amalexander.math@gmail.com} with inquiries.}

\author{William Ott$^{\dagger , 4}$}
\thanks{$\dagger$Corresponding author:
Contact \href{mailto:william.ott.math@gmail.com}{william.ott.math@gmail.com} with inquiries.}

\address{$^1$Department of Mathematics, Iowa State University, Ames, Iowa, USA}

\address{$^2$Department of Statistics, University of Kentucky, Lexington, Kentucky, USA}

\address{$^{3}$Department of Mathematics, Colgate University, Hamilton, New York, USA}

\address{$^{4}$Department of Mathematics, University of Houston, Houston, Texas, USA}


\subjclass[2020]{60, 92}
\date{\today}

\begin{abstract}
Delay is an inherent feature of genetic regulatory networks.
It represents the time required for the assembly of functional regulator proteins.
The protein production process is complex, as it includes transcription, translocation, translation, folding, and oligomerization.
Because these steps are noisy, the resulting delay associated with protein production is distributed (random).
We here consider how distributed delay impacts the dynamics of bistable genetic circuits.
We show that for a variety of genetic circuits that exhibit bistability, increasing the noise level in the delay distribution dramatically stabilizes the metastable states.
By this we mean that mean residence times in the metastable states dramatically increase.

{\bfseries Relevance to Life Sciences.}
Bistable genetic regulatory networks are ubiquitous in living organisms.
Evolutionary processes seem to have tuned such networks so that they switch between metastable states when it is important to do so, but small fluctuations do not cause unwanted switching. 
Understanding how evolution has tuned the stability of biological switches is an important problem. 
In particular, such understanding can guide the design of forward-engineered synthetic bistable genetic regulatory networks. 

{\bfseries Mathematical Content.}
We use two methods to explain this stabilization phenomenon.
First, we introduce and simulate stochastic hybrid models that depend on a switching-rate parameter.
These stochastic hybrid models allow us to unfold the distributed-delay models in the sense that, in certain cases, the distributed-delay model can be viewed as a fast-switching limit of the corresponding stochastic hybrid model.
Second, we generalize the three-states model, a symbolic model of bistability, and analyze this extension.
\end{abstract}

\maketitle

\section{Introduction}

Genetic regulatory networks (GRNs) govern how cells function and are therefore essential for unicellular and multicellular life. 
They work by utilizing molecular components as signaling agents. 
DNA codes for proteins known as transcription factors that up- or down-regulate genes in the network. 
Evolution has tuned naturally-occurring GRNs to efficiently perform a variety of tasks, ranging from morphogenesis to maintaining circadian rhythms. 
Synthetic biologists endeavor to forward-engineer GRNs in single cells, and also consortial systems that can perform complex computations \cite{chen2015emergent}. 
In a synthetic microbial consortium, the circuit is divided between several strains. 
The strains use diffusible signaling molecules to communiate with one another, thereby allowing the overall circuit to function as designed. 

\begin{figure}[t]
\centering
\includegraphics[width=.99\textwidth]{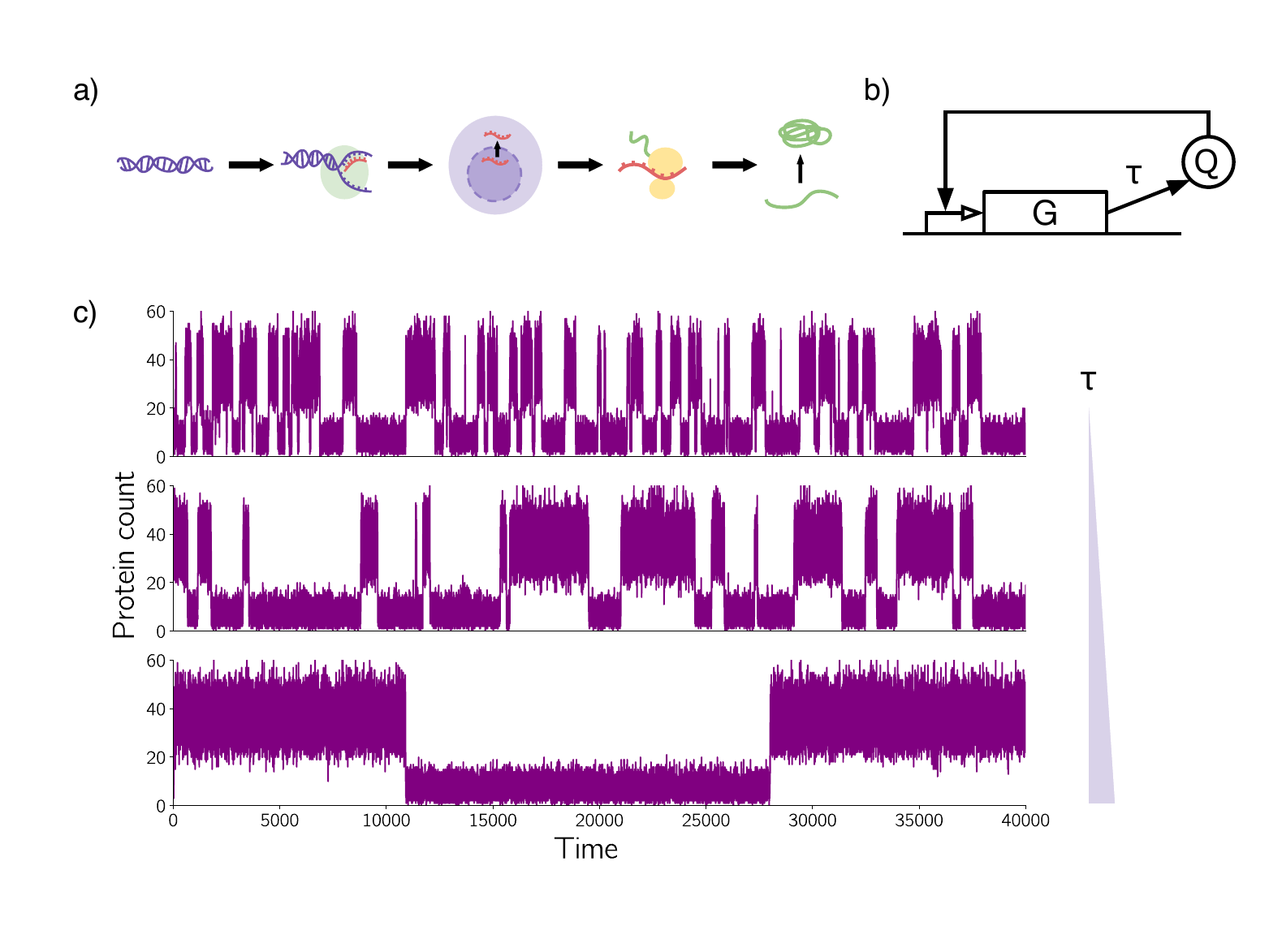}
\caption{
\textbf{(a)}
The production of transcription factor protein requires a complex sequence of noisy steps, including transcription, RNA translocation, translation, and post-translational processes such as folding and oligomerization.
\textbf{(b)}
A single-gene positive feedback loop, where the gene, $G$, produces a protein, $Q$, that upregulates its own production.
The delay, $\tau$, represents delayed protein production due to the assembly steps shown in (a).
\textbf{(c)}
Stochastic trajectories produced by the single-gene positive feedback loop.
Here, delay $\tau$ takes a fixed value.
As $\tau$ increases, hops between the two metastable states become more rare.}
\label{fig:backstory}
\end{figure}

From a mathematical point of view, the dynamics of GRNs are challenging to analyze because of the interplay between noise and delay \cite{singh2021interplay}. 
GRNs are inherently stochastic because of intrinsic biochemical reaction noise.
Delay results from the lengthy sequence of steps that must be completed in order to produce functional regulator proteins.
This sequence can include transcription, RNA translocation, translation, and post-translational processes such as folding and oligomerization \cite{Josic2011}.
See Figure~\ref{fig:backstory}a for an illustration.
We call this delay protein production delay (or transcriptional delay).

Protein production delay has two properties that should be taken into account when building predictive models for GRNs. 
First, protein production delay timescales are long enough to meaningfully impact dynamics. 
In synthetically-engineered \revision{\textit{Escherichia coli}}, for example, transcription factors require minutes to produce \cite{cheng2017timing, choi2020bayesian}. 
Second, protein production delay is distributed (random). 
This is because the assembly steps depicted in Figure~\ref{fig:backstory}a are stochastic. 

Our paper focuses on the following question. 
How does distributed delay impact the dynamics of bistable GRNs? 
Bistable GRNs admit two (or more) metastable states. 
A trajectory in such a system will tend to stay near a given metastable state for a long period of time once the trajectory enters a neighborhood of the metastable state. 
Intrinsic biochemical reaction noise (and external influences) can cause rare hops from one metastable state to another. 
Bistable GRNs (biological switches) are ubiquitous in living organisms. 
For instance, they are important for cell fate decisions during embryonic development \cite{Saez2022,Mohammed2017} and maintenance of the lac operon \cite{Zander2017,Santilln2007}. 

We show in this paper that for a variety of bistable GRNs, increasing the noise level in the delay distribution dramatically stabilizes the metastable states. 
By this we mean that mean residence times in the metastable states dramatically increase. 
In particular, we exhibit this noise-induced stabilization phenomenon for the lysis-lysogeny switch of the $\lambda$-bacteriophage, a paradigm for developmental genetic networks. 
We view stabilization of metastable states as a constructive effect, so noise (in the delay distribution) acts constructively for the systems we here consider. 
Our results are therefore consistent with a rich body of work on the benefits of noise in physical systems, a body of work that originates with early work on stochastic resonance. 

Using mathematical models to unlock understanding of GRN dynamics has been quite successful. 
This is so because such mathematical models can succinctly encode essential GRN structures. 
Indeed, it is well known that simple structures in GRNs can produce rich dynamics \cite{tyson2003sniffers, shis2018dynamics}. 
GRNs containing negative feedback loops can exhibit oscillatory dynamics \cite{elowitz2000synthetic, mather2009delay}, and GRNs containing positive feedback or mutual inhibition can exhibit bistability \cite{becskei2001positive, gardner2000construction}. 
Further, using mathematical models allows one to leverage the rich mathematical theory on delay dynamical systems. 
This theory involves oscillatory dynamics \cite{Ryzowicz2024, li2021stability}, bistability \cite{qiu2020molecular}, and chaos \cite{karamched2021delay, zeng2017stochastic, suzuki2016periodic}. 

Distributed delay has been shown to constructively impact the dynamics of GRNs in other contexts. 
It can accelerate signaling in feed-forward architectures \cite{Josic2011}, and denoise biochemical oscillators that utilize delayed negative feedback \cite{Song_Campbell2024}. 
This denoising result is surprising because precise repression timing is thought to be crucial for circadian rhythms \cite{Chae2023}, and one might conjecture that noise in the delayed negative feedback disrupts repression timing. 
See \cite{feng2016non, zhang2017emergent, kirunda2021effects} for more results on how stochastic delay influences dynamics. 

Our current paper stands most directly upon the shoulders of two works. 
Gupta et al. \cite{Gupta-2013-Transcriptional} have shown that fixed delay dramatically stabilizes a variety of bistable GRNs. 
This means that when delay is modeled by a fixed value, increasing this value produces dramatically longer mean residence times in the metastable states. 
Figure~\ref{fig:backstory}bc illustrates this effect for a single-gene positive feedback loop. 
The gene $G$ produces a protein $Q$ that requires fixed time $\tau$ to synthesize. 
The mature protein then up-regulates its own production (Figure~\ref{fig:backstory}b). 
As $\tau$ increases, hops between the low and high metastable states become rarer (Figure~\ref{fig:backstory}c). 
Kyrychko and \revision{Schwartz} \cite{Kyrychko2018} have studied the impact of distributed delay on a certain stochastic differential equation that features one metastable sink and one saddle point. 
They conclude their paper by asking about the impact of distributed delay on bistable systems. 
Our current paper responds to this question. 

We use two methods to explain how distributed delay stabilizes a variety of bistable GRNs. 
First, we introduce and simulate stochastic hybrid models that depend on a switching-rate parameter. 
Here, the biochemical reaction network is coupled to a continuous-time Markov chain. 
The Markov chain switches at a characteristic rate set by the switching-rate parameter. 
Its states correspond to different fixed delay values. 
These stochastic hybrid models allow us to unfold the distributed-delay models in the sense that, in certain cases, the distributed-delay model can be viewed as a fast-switching limit of the corresponding stochastic hybrid model. 
The key question is then how the dynamics depend on the switching-rate parameter. 
The idea to switch between different fixed delay values is inspired by \cite{Karamched_Miles_2023}. 
Second, we generalize the three-states model from \cite{Gupta-2013-Transcriptional}, a symbolic model of bistability originally designed to explain why fixed delay dramatically stabilizes a variety of bistable GRNs. 
Our analysis of this extension explains why the stochastic hybrid model behaves as it does when the switching rate is low.

\section{Increasing delay variability stabilizes metastable states}

We show that for three bistable GRNs, increasing delay variability stabilizes the metastable states.
The models are the co-repressive toggle switch, the single-gene positive feedback loop, and a model of the lysis-lysogeny switch of the $\lambda$-bacteriophage.
More precisely, we hold the mean of the delay distribution fixed and increase the coefficient of variation (CV).
As delay CV increases, mean residence times in \revision{the} metastable states respond unimodally, meaning that they first dramatically increase before eventually tapering off.
We show that this stabilization result is robust with respect to choice of delay distribution.
\revision{Our simulations have utilized ranges for delay mean and delay CV such that the delay timescale is on the order of protein half-life and much shorter than mean residence times in the metstable states.
This delay timescale is biophysically appropriate for bistable genetic regulatory networks.}

\subsection{Modeling hierarchy}

This paper utilizes the modeling hierarchy for biochemical reaction networks that is described and analyzed in \cite{gupta2014modeling}.
The framework allows for a combination of instantaneous reactions and delayed reactions.
The first layer of the hierarchy is \revision{the delay birth-death (dBD) process, a} stochastic process that tracks molecule counts \cite{Schlicht-2008-delay}.
This layer models GRNs that operate at low molecule counts.
Each chemical reaction has an associated propensity and state-change vector.
If a given reaction is delayed, it comes equipped with a Borel probability measure (the delay distribution).
Each time this delayed reaction initiates, a sample is drawn from the delay distribution to give the completion time for this particular instantiation of the delayed reaction.
If a given reaction is instantaneous, it completes right when it fires.

Our stochastic simulations sample from this stochastic process.
To generate these samples, we utilize a delay stochastic simulation algorithm (dSSA).
The dSSA is a delay variant of the Gillespie algorithm \cite{Gillespie1976}.
We review the dSSA in Appendix~\ref{dSSAalg}.

The second layer of the hierarchy is relevant when molecule counts are moderate.
At this layer, the evolution of the biochemical reaction network \revision{is} described by an integro-delay stochastic differential equation known as the delay chemical Langevin equation (dCLE).
The dCLE is expressed in terms of molecule concentrations.
Importantly, its fluctuation-dissipation relation is such that both the drift and diffusion involve delay.
The third layer of the hierarchy is obtained in the limit of large system size.
Here, molecule concentrations evolve according to a deterministic integro-delay differential equation known as the delay reaction rate equation (dRRE).

\subsection{The primary experiment}

For the bistable systems outlined below, we have performed the following experiment: (1) choose a delay distribution and fix the mean; (2) for a range of delay CVs, simulate the system using the dSSA until sufficiently many transitions between metastable states occur; (3) collect residence times in each of the metastable states.
\revision{Notice that this experiment can be performed for the dBD representation of the bistable GRN or the dCLE representation of the bistable GRN.
We have opted for the former in this paper because our focus here is on rare-event statistics when molecule counts are low.}

\revision{Following \cite{Gupta-2013-Transcriptional}, we} utilize the thresholding technique \revision{explained below} to measure residence times. 
Let $X(t)\in \mathbb{Z}^J$ represent the number of proteins of $J$ species at time $t$, and suppose the corresponding dRRE has two stable steady states at protein concentration vectors $x_{s0}$ and $x_{s1}$.
We define neighborhoods $U_0$ of $x_{s0}$ and $U_1$ of $x_{s1}$.
Suppose that $X(0) \in U_{0}$.
The $n^{\rm{th}}$ time at which the system transitions from one metastable state to the other, denoted $G_n$, is defined as $G_n =\inf \{ t>G_{n-1} : X(t) \in U_i \}$, where $i=1$ if the trajectory was most recently in $U_0$ and $i=0$ if the trajectory was most recently in $U_1$.
Set $G_0=0$.
The residence times in $U_{0}$ and $U_{1}$ are defined by $R^{n}_{s0} = G_{n+1}-G_{n}$ for $n$ even and $R^{n}_{s1} = G_{n+1} - G_{n}$ for $n$ odd.
Intuitively, a residence time for $U_{0}$ may be thought of as follows.
Once the trajectory enters $U_{0}$, its residence-time clock starts.
The trajectory is then free to leave and reenter $U_{0}$ multiple times; the residence-time clock only stops when the trajectory first enters $U_{1}$. \revision{The sizes of the neighborhoods $U_0$ and $U_1$ are hyperparameters that we choose based on the particular bistable system.
Our computational results are not sensitive to these neighborhood sizes, as long as each neighborhood is small enough to ensure that typical residence times in each neighborhood are long compared to protein half-life.
In other words, the neighborhoods should be small enough to ensure that we are in the biophysically relevant regime.
We report our chosen neighborhood sizes for each system we simulate in the relevant figure caption.}

\subsection{Noisy delay stabilizes the co-repressive toggle switch}\label{co-repressive toggle}

\begin{figure}[ht]
\centering
\includegraphics[width=.99\textwidth]{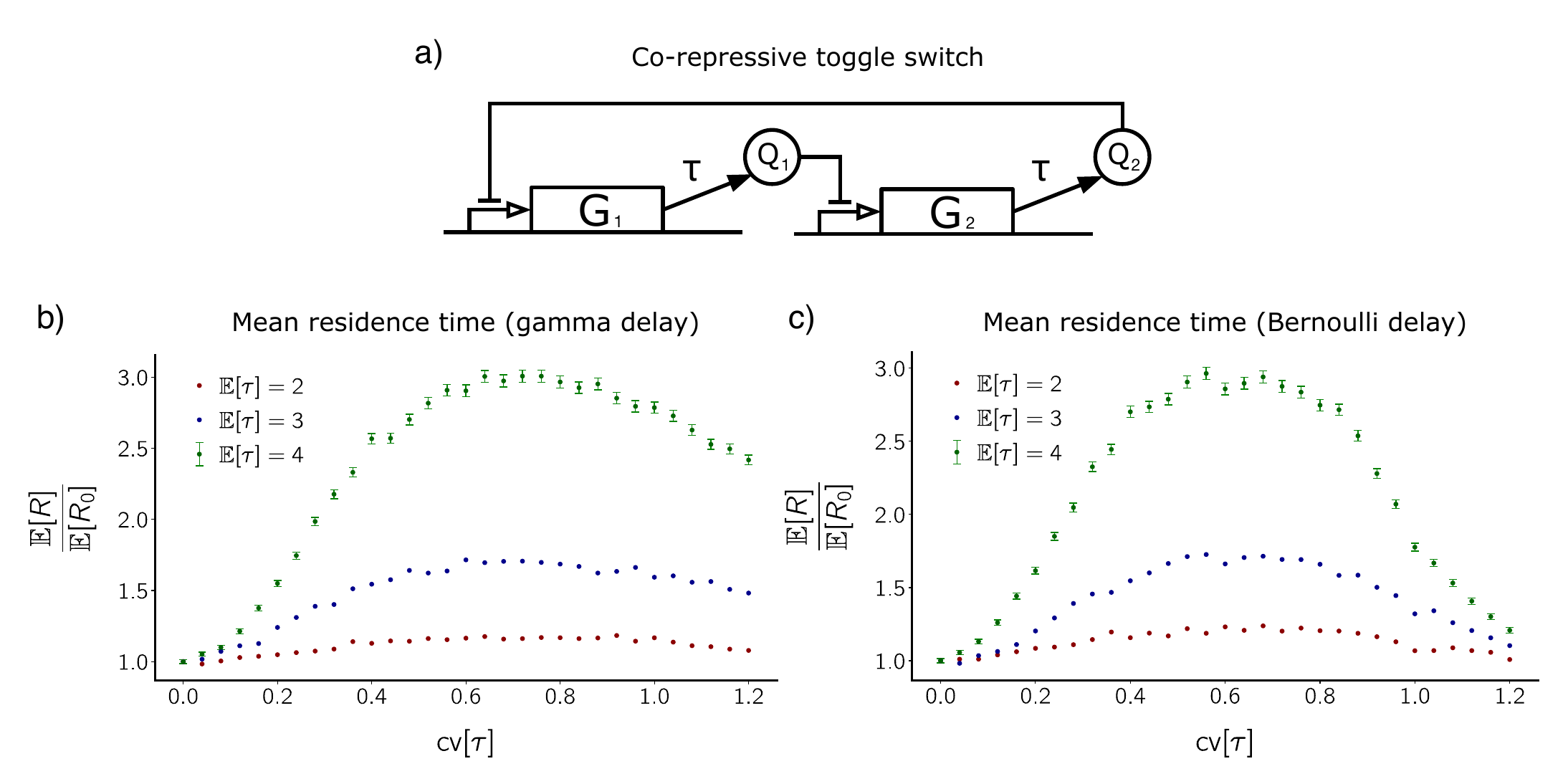}
\caption{\textbf{Distributed delay stabilizes the co-repressive toggle switch.}
\textbf{(a)} Circuit diagram for the co-repressive toggle switch.
\textbf{(b)} Mean residence time, $\E[R]$, for the metastable states of the co-repressive toggle switch, plotted against coefficient of variation of gamma-distributed delay.
Color corresponds to each of three values of delay mean.
Since the system is symmetric, residence times have been pooled over the two metastable states.
$\E[R]$ is normalized by $\E[R_0]=1.93\times10^4$, $7.21\times10^4$, and $2.15\times10^5$ for $\E[\tau]=2$ (red), $3$ (blue), and $4$ (green), respectively. Error bars for $\E[\tau]=2$ and $\E[\tau]=3$ curves are approximately the size of the points. \revision{Residence times were measured using neighborhoods defined as closed balls of radius 5 around the metastable states of $(N_{Q_1}, N_{Q_2})=(1.53, 30.11)$ and $(N_{Q_1}, N_{Q_2})=(30.11, 1.53)$.}
\textbf{(c)}
Analogous to (b), but the delay distribution is Bernoulli (symmetric about its mean). 
}
\label{fig:toggle}
\end{figure}

The first system we have simulated is the co-repressive toggle switch, as this GRN is well-studied and known to exhibit bistability. Such a toggle switch can be found in the Cyanobacteria circadian oscillator \cite{Ishiura_Kutsuna_Aoki_Iwasaki_Andersson_Tanabe_Golden_Johnson_Kondo_1998a}.
The network involves two genes whose products repress the production of each other, as depicted in Figure~\ref{fig:toggle}a.
The reactions consist of protein production (delayed) and dilution (instantaneous).
The system of dRREs for the co-repressive toggle switch is given by
\begin{subequations}\label{eqn:Bistab_Corepressive}
    \begin{align}
        \dot{q_1}(t)&=\int_0^\infty\frac{\beta}{1+q_2(t-\tau)^2/\kappa}\,d\Prob(\tau)-\gamma q_1(t)\\
        \dot{q_2}(t)&=\int_0^\infty\frac{\beta}{1+q_1(t-\tau)^2/\kappa}\,d\Prob(\tau)-\gamma q_2(t),
    \end{align}
\end{subequations}
where $q_1$ and $q_2$ represent protein concentrations of \revision{the proteins $Q_{1}$ and $Q_{2}$}, respectively.
The parameters for this system are interpreted symmetrically as follows: $\beta$ is the maximal protein production rate, $\kappa^2$ produces half-maximal protein production rate, $\gamma$ is the dilution rate, and $\Prob$ is a Borel probability measure that describes the distribution of delay (the delay kernel).

For our stochastic simulations, we have chosen parameters for which there exist two metastable states, \revision{($N_{Q_{1}}$ high, $N_{Q_{2}}$ low) and ($N_{Q_{1}}$ low, $N_{Q_{2}}$ high).
Here, $N_{Q_{1}}$ and $N_{Q_{2}}$ denote protein counts for the proteins $Q_{1}$ and $Q_{2}$,} respectively.
See Table~\ref{tab:Bistab_corepressive} for reaction propensities and parameter choices.

\begin{table}[ht]
    \centering
    \caption{Propensities and parameter values for the co-repressive toggle switch~\cite{Gupta-2013-Transcriptional}. 
    Bold font denotes reactions with delay.}
    \renewcommand{\arraystretch}{1.4} 

    \begin{tabular}{@{\hspace{2em}} c @{\hspace{8em}}c @{\hspace{8em}} c @{\hspace{2em}}}
    \toprule
    \textrm{Reaction}&
    \textrm{Propensity}&
    \textrm{Parameters}\\
    \midrule
    $\boldsymbol{\emptyset\rightarrow}\mathbf{Q_1}$ & $\frac{\beta \kappa}{\kappa+\revision{N_{Q_2}}^2}$ & $\beta = 31.6404\log(2),\ \kappa=\frac{1000}{21.6404}$\\ \hline
    $\boldsymbol{\emptyset\rightarrow}\mathbf{Q_2}$ & $\frac{\beta \kappa}{\kappa+\revision{N_{Q_1}}^2}$ & $\beta = 31.6404\log(2),\ \kappa=\frac{1000}{21.6404}$\\ \hline
    $Q_1\rightarrow\emptyset$ & $\gamma \revision{N_{Q_1}}$ & $\gamma=\log(2)$\\ \hline
    $Q_2\rightarrow\emptyset$ & $\gamma \revision{N_{Q_2}}$ & $\gamma=\log(2)$\\
    \bottomrule
    \end{tabular}
    \label{tab:Bistab_corepressive}
    \end{table}

Figure~\ref{fig:toggle}b shows mean residence time $\ev [R]$ for the co-repressive toggle switch as a function of delay CV (with delay mean fixed).
Here, the delay follows a gamma distribution with means $2$, $3$, and $4$.
Because the system is symmetric, we pool residence times over the two metastable states.
For each delay mean, we normalize mean residence time by the expected residence time when delay CV is zero (fixed delay), denoted $\ev [R_{0}]$. Note that $\ev [R_{0}]$ increases as a function of fixed delay value \cite{Gupta-2013-Transcriptional}. As $\cv [\tau ]$ increases starting from zero, normalized mean residence time substantially increases (before eventually peaking and then decreasing).
Perhaps counterintuitively, injecting noise into the delay distribution stabilizes the metastable states.
Noise is therefore acting constructively here if such stabilization is desirable.

Figure~\ref{fig:toggle}c shows mean residence time $\ev [R]$ for the co-repressive toggle switch as a function of delay CV (with delay mean fixed), but this time the delay follows a Bernoulli distribution.
Here, the Bernoulli distribution is symmetric about delay means $2$, $3$, and $4$.
Once again, we pool residence times over the two metastable states.
As with the gamma distribution, we see that normalized mean residence time substantially increases as $\cv [\tau]$ increases, over a range of values of $\cv [\tau ]$.
This shows that the stabilization effect is robust with respect to choice of delay distribution.

\subsection{Noisy delay stabilizes a variety of bistable GRNs}
\label{sSection:robustness}

To test the robustness of the stabilizing effect of distributed delay, we have simulated two more bistable systems.
These are the single-gene positive feedback loop and a model of the $\lambda$-bacteriophage switch.
Throughout Section~\ref{sSection:robustness}, delay is gamma-distributed.

\subsubsection{Single-gene positive feedback loop}

\begin{figure}[ht]
\centering
\includegraphics[width=.99\textwidth]{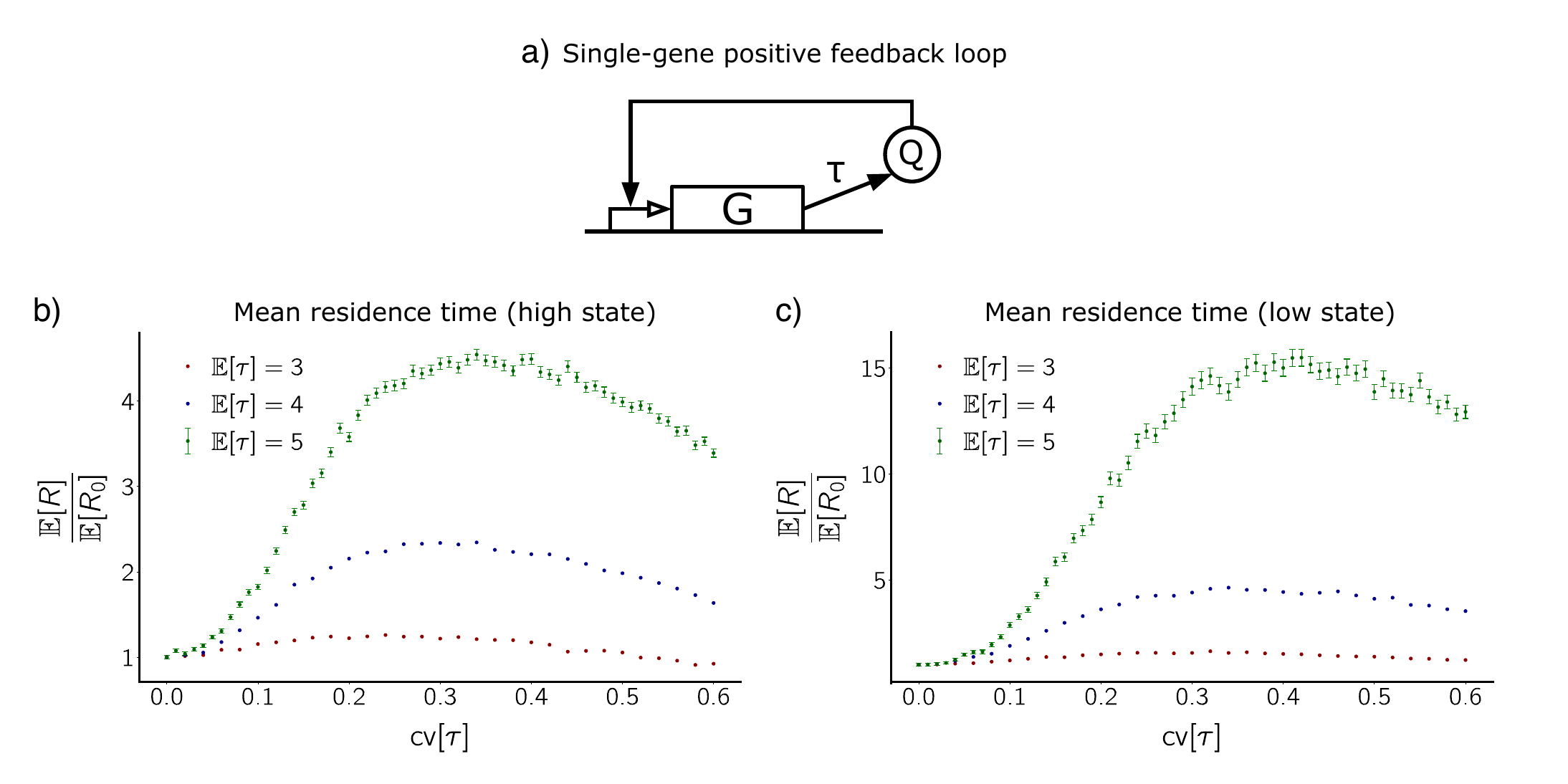}
\caption{\textbf{Distributed delay stabilizes the single-gene positive feedback loop.}
\textbf{(a)}
Circuit diagram for the single-gene positive feedback loop.
\textbf{(b), (c)}
Mean residence time in the high metastable state (b) and the low metastable state (c) for the single-gene positive feedback loop as a function of the CV of gamma-distributed delay.
Mean residence time is normalized by the mean residence time at CV zero for each of the three delay means.
The normalizations are $\E[R_0]=1.82\times10^4$, $1.50\times10^4$, and $7.16\times10^3$ for the low state and $\E[R_0]=6.69\times10^4$, $1.61\times10^5$, and $2.48\times10^5$ for the high state, with delay means $\E[\tau]=3$ (red), $4$ (blue), and $5$ (green), respectively. Error bars for $\E[\tau]=3$ and $\E[\tau]=4$ curves are approximately the size of the points. \revision{Residence times were measured using neighborhood $[0,8.4)$  for metastable state of $N_{Q}=7.2$  and neighborhood $(30,\infty)$ for metastable state of $N_{Q}=36$.}
}
\label{fig:single_gene}
\end{figure}

This GRN involves a single gene whose product increases the rate of its own production, as shown in Figure~\ref{fig:single_gene}a.
The dRRE for this system is given by
\begin{equation}\label{eqn:Bistab_single-speciesMFD}
\dot{q}(t)=\int_0^\infty\alpha+\beta\frac{q(t-\tau)^b}{c^b+q(t-\tau)^b}\,d\Prob(\tau)-\gamma q(t),
\end{equation}
where $q$ represents protein concentration.
The parameters can be interpreted as follows: $\alpha$ represents the basal production rate while $\alpha+\beta$ is the maximal production rate, $c$ corresponds to production rate $\alpha+\frac\beta2$, $b$ is the Hill coefficient for the activation function, and $\gamma$ is the dilution rate.  

For our stochastic simulations, we have selected parameters for which there exist two metastable states, \revision{$N_Q$} low and \revision{$N_Q$} high.
Here, \revision{$N_Q$} represents protein count.
See Table~\ref{tab:Bistab_single-species} for reaction propensities and parameter values.

\begin{table}[ht]
\centering
\caption{Propensities and parameter values for the single-gene positive feedback loop \cite{Gupta-2013-Transcriptional}. 
Bold font denotes the reaction with delay.}
    \renewcommand{\arraystretch}{1.4} 

    \begin{tabular}{@{\hspace{2em}} c @{\hspace{8em}}c @{\hspace{8em}} c @{\hspace{2em}}}
    \toprule
    \textrm{Reaction}&
    \textrm{Propensity}&
    \textrm{Parameters}\\
    \midrule
    $\boldsymbol{\emptyset \rightarrow}\mathbf{Q}$ & $\alpha+\beta \frac{\revision{N_Q}^b}{c^b + \revision{N_Q}^b}$ & $\alpha = 5,\ \beta = 20,\ b=10,\ c = 19$\\ \hline
    $Q \rightarrow \emptyset$ & $\gamma \revision{N_Q}$ & $\gamma = \log(2)$\\
    \bottomrule
    \end{tabular}
    \label{tab:Bistab_single-species}
\end{table}

Figure~\ref{fig:single_gene} shows mean residence time as function of delay CV for three values of delay mean.
Once again, we normalize by the mean residence time when delay CV is zero (fixed delay).
Since the single-gene positive feedback loop is not symmetric, we report the data for the low state and the high state separately.
Observe that the qualitative behavior of the normalized mean residence time curves for the single-gene positive feedback loop is consistent with that of the co-repressive toggle switch.
This indicates robustness of the noise-induced stabilization effect with respect to system architecture.

\subsubsection{A bacteriophage switch}

\begin{figure}[ht]
\centering
\includegraphics[width=.99\textwidth]{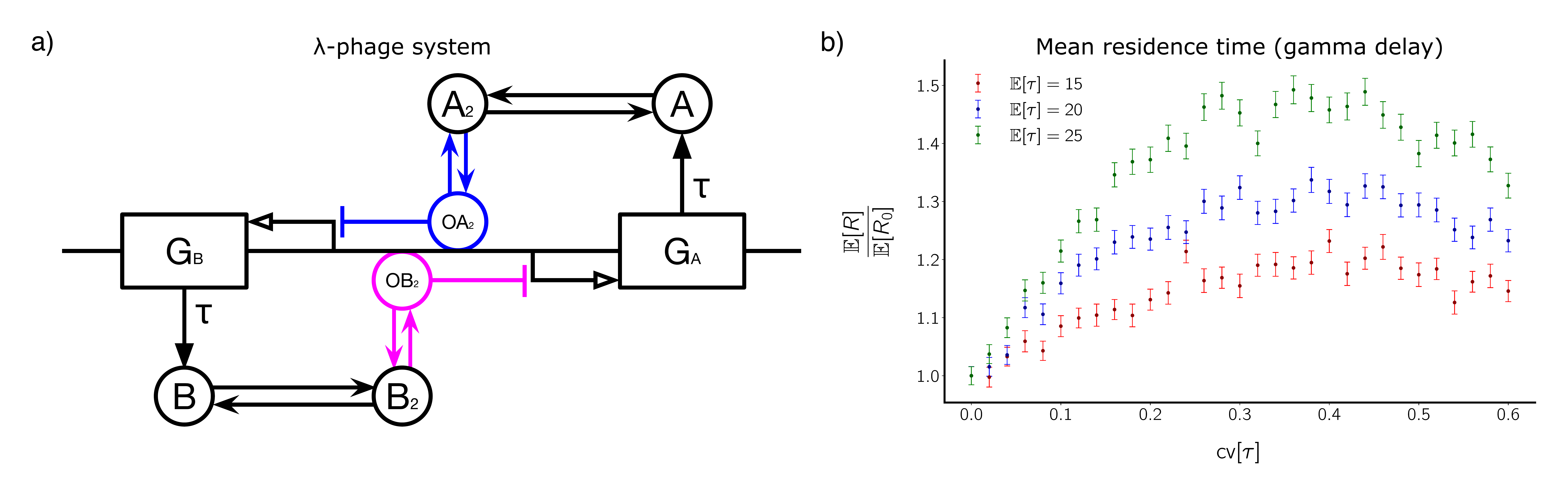}
\caption{\textbf{Distributed delay stabilizes the $\lambda$-bacteriophage model.}
\textbf{(a)}
Circuit diagram for the $\lambda$-bacteriophage model.
Colored parts of the diagram represent competing reactions; if for instance, if $A_2$ binds to $O$ to form $OA_2$, then $B_2$ cannot bind to $O$.
\textbf{(b)}
Mean residence time $\ev [R]$ in the metastable states for the $\lambda$-bacteriophage model plotted against coefficient of variation of gamma-distributed delay.
$\ev[R]$ is normalized by $\ev[R_0]=7.38\times10^4$, $1.41\times10^5$, and $2.60\times10^5$ for $\ev[\tau]=15$ (red), $20$ (blue), and $25$ (green), respectively.
\revision{Residence times were measured using neighborhoods defined as closed balls of radius 5 around the metastable states of $(A_T, B_T)=(0.07, 27.50)$ and $(A_T, B_T)=(27.50, 0.07)$.}
}
\label{fig:phage}
\end{figure}

The $\lambda$-bacteriophage is a naturally occurring virus that infects \textit{E. coli} cells. Its dynamics include a bistable switch between a dormant (\revision{lysogenic}) state and an active (lytic) state.
We do not model the system in full biological detail.
Rather, we utilize a model of the $\lambda$-bacteriophage from \cite{warren2005chemical} that captures the essential reactions leading to bistability \cite{cherry2000make}.
Throughout our discussion of the $\lambda$-bacteriophage, we slightly abuse notation by using the same notation for species name and molecule count of that species.
The system has two transcription factors (TFs), denoted $A$ and $B$, and the genes coding for these TFs are adjacent and share an operator site in the DNA. The two TFs dimerize to form $A_2$ and $B_2$, and these dimers compete to bind to the operator site, $O$. Binding of $A_2$ to $O$ blocks transcription of $B$, and binding of $B_2$ to $O$ blocks transcription of $A$. We denote the dimer:operator complexes as $OA_2$ and $OB_2$, respectively. These interactions are depicted in Figure~\ref{fig:phage}a, with the competing sets of reactions depicted in different colors. Production of $A$ and $B$ is subject to transcriptional delay, but the other reactions are instantaneous.

The biochemical reactions of the model are summarized in Table~\ref{tab:Phage}.
The first four reactions denote monomer expression from either the bound or unbound operator site, followed by degradation of $A$ and $B$ monomer, and then four reactions for the assembly or disassembly of dimers, and finally four reactions for the dimers binding to and unbinding from the operator.
Warren and ten Wolde 2005 \cite{warren2005chemical} conduct a mean-field analysis of this system assuming that there is one operator site, and show that the system is bistable for biologically reasonable parameter values.
For our stochastic simulations, we use parameter values within this range.

\begin{table}[ht]
\centering
\caption{Reactions and propensities for the $\lambda$-bacteriophage model.
Bold font is used for reactions with delay.
The parameter values we use for stochastic simulations are as in \cite{Gupta-2013-Transcriptional}: $k_b=k_f=k_{\rm{on}}=5$, $k_A=k_B=k_{\rm{off}}=1$, $\mu_A=\mu_B=.3$.
}
\renewcommand{\arraystretch}{1.4} 
\begin{tabular}{@{\hspace{2em}} c @{\hspace{8em}}c  @{\hspace{2em}}}
    \toprule
    \textrm{Reaction}&
    \textrm{Propensity}\\
    \midrule
    $\boldsymbol{O\rightarrow O+A}$ &$k_A\times O$  \\\hline
    $\boldsymbol{OA_2\rightarrow OA_2+A}$ &$k_A\times OA_2$  \\\hline
    $\boldsymbol{O\rightarrow O+B}$ &$k_B\times O$  \\\hline
    $\boldsymbol{OB_2 \rightarrow OB_2+B}$ &$k_B\times OB_2$  \\\hline
    $A\rightarrow \emptyset$ &$\mu_A\times A$  \\\hline
    $B\rightarrow \emptyset$ &$\mu_B\times B$   \\\hline
    $A+A\rightarrow A_2$ &$k_f\times \frac{A(A-1)}2$   \\\hline
    $A_2\rightarrow A+A$ &$k_b\times A_2$   \\\hline
    $B+B \rightarrow B_2$ &$k_f\times \frac{B(B-1)}2$   \\\hline
    $B_2\rightarrow B+B$&$k_b\times B_2$   \\\hline
    $O+A_2\rightarrow OA_2$ &$k_{\rm{on}}\times O\times  A_2$   \\\hline
    $OA_2\rightarrow O+A_2$ &$k_{\rm{off}}\times OA_2$   \\\hline
    $O+B_2\rightarrow OB_2$&$k_{\rm{on}}\times O\times  B_2$   \\\hline
    $OB_2\rightarrow O+B_2$ &$k_{\rm{off}}\times OB_2$   \\
    \bottomrule
\end{tabular}
\label{tab:Phage}
\end{table}

This system contains seven molecular species: $A$, $B$, $A_2$, $B_2$, $O$, $OA_2$, and $OB_2$. To observe bistability, it is natural to project the dynamics onto a two-dimensional subspace spanned by $A_T=A+2A_2+2OA_2$ and $B_T=B+2B_2+2OB_2$.
We project in this way and then examine switching between (neighborhoods of) two metastable states, ($A_{T}$ low, $B_{T}$ high) and ($A_{T}$ high, $B_{T}$ low).

Figure~\ref{fig:phage}b shows mean residence time as a function of $\cv [\tau ]$ when the delay follows a gamma distribution.
Because the $\lambda$-bacteriophage model is symmetric, we pool residence times from both metastable states to produce the means.
We normalize mean residence time by mean residence time when delay CV is zero (fixed delay).
As we have seen with the co-repressive toggle switch and the single-gene positive feedback loop, normalized mean residence time exhibits a unimodal response for the $\lambda$-bacteriophage model.
As $\cv [\tau ]$ increases, normalized mean residence time first increases, then plateaus, and finally decreases.

\revision{Residence-time statistics for the lysogenic state have been measured experimentally~\cite{Zong_2010_Lysogen}.
This experimental paper reports that lysogen stability exhibits a simple exponential dependence on
the frequency of activity bursts from the fate-determining gene.
This finding is consistent with our computational results for the $\lambda$-bacteriophage.
See~\cite{Lei_2015_Biological} for an examination of the sources of intrinsic and extrinsic noise in the $\lambda$-bacteriophage system.}

Overall, we have shown that noisy delay stabilizes three bistable GRNs and that this effect appears to be robust with respect to choice of delay distribution.
In the following section, we propose an explanation for this stabilization phenomenon that is not model-specific.

\section{Explaining noise-induced stabilization by unfolding the Bernoulli distribution}
\label{sec:unfolding}

The result that injecting noise into the delay distribution stabilizes metastable states in stochastic GRNs is surprising.
We introduce a stochastic hybrid modeling framework to explain this phenomenon.

\subsection{An intuitive description of unfolding}

We present our stochastic hybrid modeling framework in general in Appendix~\ref{sec:stochastic_hybrid}.
For Section~\ref{sec:unfolding}, it suffices to intuitively describe how we use a special case of the framework to explain the stabilization phenomenon we have discovered.
We now give this intuitive description.

Consider the co-repressive toggle switch with Bernoulli delay from Figure~\ref{fig:toggle}.
Write the delay measure as $(1/2) \delta_{\mu - \sigma} + (1/2) \delta_{\mu + \sigma}$, where $\delta$ denotes the Dirac-$\delta$ measure, $\mu$ is the mean of the Bernoulli distribution, and $\sigma$ is a parameter.
When simulating this system with the dSSA, the production reactions are handled as follows.
Each time a production reaction initiates, a delay time \textit{for that particular reaction} is selected from the Bernoulli distribution.

We can unfold this system in the following way.
Run a two-state continuous-time Markov chain in the background.
Parametrize this Markov chain by $r$, where $r$ gives the switching rate from state $0$ to state $1$ and from state $1$ to state $0$.
When the Markov chain is in state $0$, every production reaction that initiates uses delay value $\mu - \sigma$.
On the other hand, when the Markov chain is in state $1$, every production reaction that initiates uses delay value $\mu + \sigma$.
By doing this, we have created a family of stochastic hybrid systems parameterized by $r$, the switching rate.
The co-repressive toggle switch with Bernoulli delay then arises as the fast-switching ($r \to \infty$) limit of this family.

We now study the unfolding of the co-repressive toggle switch with Bernoulli delay in order to uncover the mechanisms behind the stabilization phenomenon that we have discovered.

\subsection{How switching rate tunes the behavior of the unfolded co-repressive toggle switch}

\begin{figure}[ht]
\includegraphics[width=.99\textwidth]{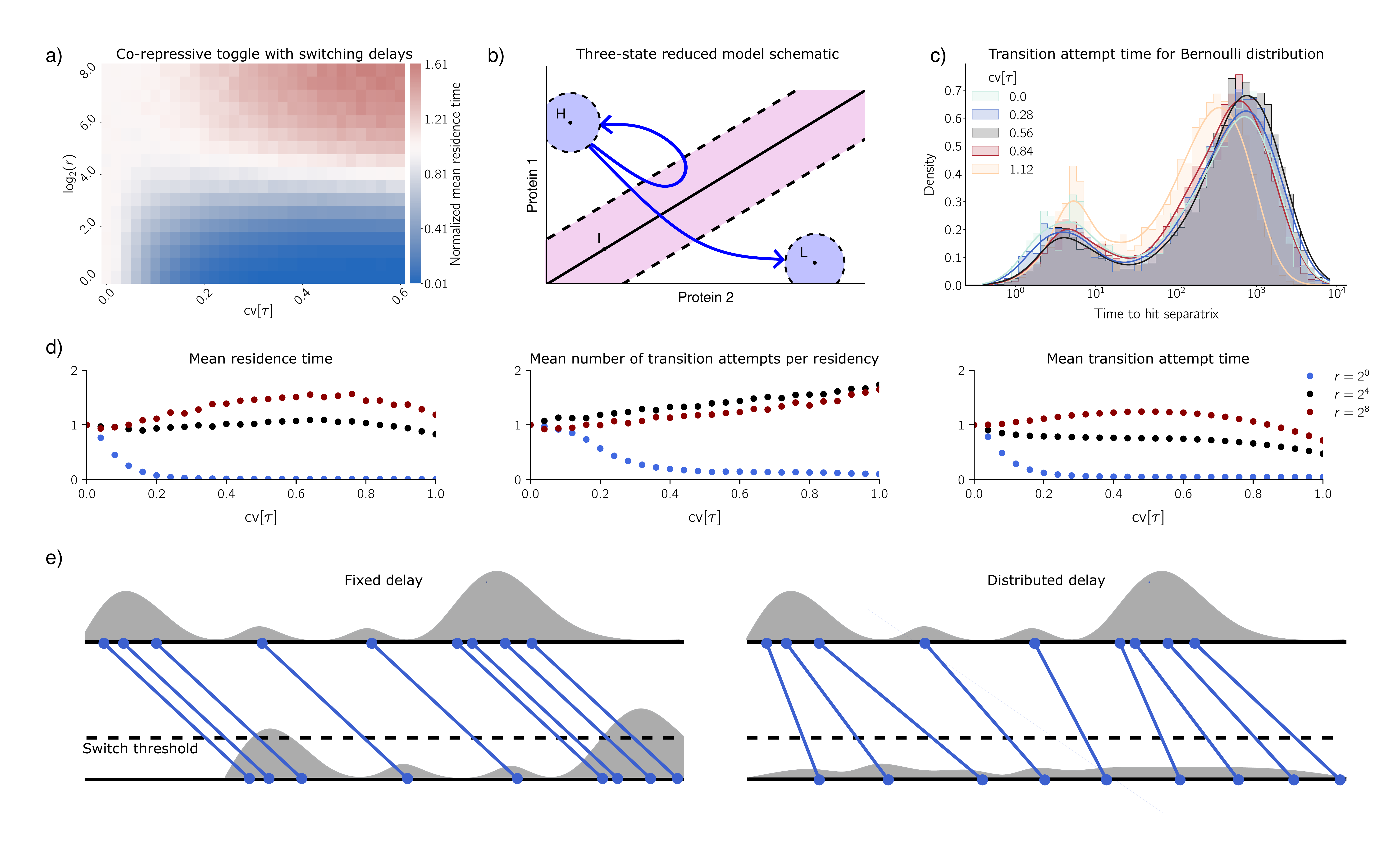}
\caption{
\textbf{Unfolding of the co-repressive toggle switch.}
\textbf{(a)}
Mean residence time (heat) as a function of switching rate $r$ and $\cv [\tau ]$.
Here the delay mean is $3$.
Mean residence time is normalized by its value at $\cv [\tau ] = 0$.
\textbf{(b)}
A symbolic coding of the co-repressive toggle switch.
Here $H$ and $L$ are neighborhoods of the two metastable states, and $I$ is a thickening of the line of symmetry.
Blue arrows indicate a failed transition ($H \rightarrow I \rightarrow H$) and a successful transition ($H \rightarrow I \rightarrow L$).
\textbf{(c)}
Histograms of transition time for the $H \rightarrow I$ or $L \rightarrow I$ transition. \revision{Transition times were measured using neighborhoods defined as closed balls of radius 6 around the metastable states, and a band of width 12 centered on the separatrix.}
For (c), the system is the co-repressive toggle switch with Bernoulli-distributed delay.
\textbf{(d)}
Timing means for the unfolded co-repressive toggle switch for three different values of the switching rate.
\textbf{(e)}
Schematic illustrating that distributed delay breaks apart Poissonian bursts of protein production activity.
Top: initiation of protein production.
Bottom: completion of protein production.
Left: fixed delay.
Right: distributed delay.
}
\label{fig:unfolding}
\end{figure}

We have simulated the unfolded co-repressive toggle switch to collect residence time data.
Figure~\ref{fig:unfolding}a shows mean residence time as a function of switching rate (vertical axis) and $\sigma / \mu$ (horizontal axis).
Note that $\sigma / \mu$ is the CV of the Bernoulli distribution $(1/2) \delta_{\mu - \sigma} + (1/2) \delta_{\mu + \sigma}$.
Mean residence time is normalized by mean residence time for $\cv [\tau ] = 0$.
For this heatmap, $\mu$ has been set to $\mu = 3$ to mirror Section~\ref{co-repressive toggle}.
The row of the heatmap corresponding to the fastest switching rate should therefore approximate the result in Figure~\ref{fig:toggle}c.

Horizontal slices of the heatmap tell an intriguing story.
When switching rate is low, increasing $\cv [\tau ]$ causes mean residence time to decrease (blue region), indicating destabilization.
When switching rate is high, however, increasing $\cv [\tau ]$ causes mean residence time to increase (red region), indicating stabilization.
The latter of these behaviors is consistent with the results in Section~\ref{co-repressive toggle}.
Overall, we see a transition from destabilization to stabilization as switching rate increases.
This suggests that to understand the stabilization phenomenon that occurs for the co-repressive toggle with Bernoulli delay, it would be wise to compare the switching timescale in the unfolded co-repressive toggle system with other important timescales.
We proceed along these lines.

\subsection{Explaining the heatmap using the three-states model}

We explain the heatmap using the three-states reduced model (RM) introduced by Gupta et al. in~\cite{Gupta-2013-Transcriptional}.
The idea behind this model is to represent the phase space of the genetic regulatory network symbolically.
The three states, $H$, $I$, and $L$, correspond to three pairwise-disjoint subsets of phase space.
The coding we use for the co-repressive toggle switch is shown in Figure~\ref{fig:unfolding}b.
The states $H$ and $L$ correspond to neighborhoods of the two metastable states.
State $I$ is a thickening of the line of symmetry.

The idea is now to track transitions between these three sets to learn how transition dynamics depend on switching rate and $\cv [\tau ]$.
We focus on transitioning from $H$ to $L$ (from one metastable state to another).
In order to reach state $L$ from state $H$, there will be some number of failed transition attempts (defined by transition $H \rightarrow I$ followed by $I \rightarrow H$), before a successful transition attempt (defined by transition $H \rightarrow I$ followed by $I \rightarrow L$).
We therefore estimate expected residence time in state $H$, $R_{H}$, as
\begin{equation}
\label{eq:RT_estimate}
\begin{aligned}
\ev [R_{H}] &\sim \ev [\# \text{ of } H \rightarrow I \rightarrow H] \big( \ev [\text{time for } H \rightarrow I] + \ev [\text{time for } I \rightarrow H] \big)
\\
&\qquad {}+
\big( \ev [\text{time for } H \rightarrow I] + \ev [\text{time for } I \rightarrow L] \big).
\end{aligned}
\end{equation}
The dominant terms in~\eqref{eq:RT_estimate} are the expected number of failed transitions and the expected time for the $H \rightarrow I$ transition.

We have calculated these two dominant terms via simulation.
Since the co-repressive toggle switch is symmetric, we have pooled over both metastable states when collecting data.
Figure~\ref{fig:unfolding}d shows mean residence time as a function of $\cv [\tau ]$ for three values of switching rate $r$, $r=2^{0}$ (slow switching, blue), $r=2^{4}$ (moderate switching rate, black), and $r=2^{8}$ (fast switching, red).
Mean residence time is normalized by its value at $\cv [\tau ] = 0$.
These three horizontal slices of the heatmap correspond to destabilization, a neutral response, and stabilization, respectively.
The second and third panels of Figure~\ref{fig:unfolding}d show (normalized) mean number of transition attempts per residency and (normalized) mean transition attempt time as functions of $\cv [\tau ]$ for the same three values of switching rate $r$.

When $r=2^{0}$, both mean number of transition attempts per residency and mean transition attempt time are decreasing functions of $\cv [\tau ]$.
This accounts for the observed destabilization when $r=2^{0}$, namely that mean residence time is a decreasing function of $\cv [\tau ]$ when $r=2^{0}$.

Comparing $r=2^{4}$ and $r=2^{8}$ data, these switching rates produce similar curves for mean number of transition attempts per residency versus $\cv [\tau ]$.
However, the curves for mean transition attempt time show divergent behavior.
Mean transition attempt time decreases over $\cv [\tau ] \in [0, 0.6]$ when $r=2^{4}$, but increases over (roughly) this range when $r=2^{8}$.
Together, these observations suggest that the stabilization of the metastable states that we see in the fast-switching limit -- the co-repressive toggle switch with Bernoulli delay -- happens because over a range of values of $\cv [\tau ]$, increasing $\cv [\tau ]$ causes mean transition attempt time to increase.

The probability density function for mean transition attempt time shows that, indeed, mean residence time and mean transition attempt time are correlated.
Figure~\ref{fig:unfolding}c shows how the density for transition attempt time varies with $\cv [\tau ]$ for the co-repressive toggle switch with Bernoulli delay.
As $\cv [\tau ]$ increases, we observe that mass shifts to the right, before eventually exhibiting a leftward shift.
This is consistent with the unimodal behavior of mean residence time in Figure~\ref{fig:toggle}c.
(We hypothesize that the densities for transition attempt time are bimodal because of delay-induced memory.
If a trajectory has recently entered state $H$ from state $I$, for instance, it remembers having recently been in state $I$ and therefore might quickly reenter state $I$.
This accounts for the left peak in Figure~\ref{fig:unfolding}c.
However, if a trajectory has been in state $H$ for a long time, it only remembers having been in state $H$, thereby making it harder to transition to $I$.
This accounts for the right peak in Figure~\ref{fig:unfolding}c.

We must answer two questions to finish explaining the heatmap in Figure~\ref{fig:unfolding}a.
First, when switching rate is high, why does mean transition attempt time increase with $\cv [\tau ]$ for $\cv [\tau ] \in [0, \sim0.6]$?
Second, when switching rate is low, why is mean number of transition attempts per residency a decreasing function of $\cv [\tau ]$?
The answers to these two questions explain why we see a region of stabilization and a region of destabilization, respectively.

We answer the first question by arguing that distributed delay acts convolutionally to break apart Poissonian bursts.
Consider the single-gene positive feedback loop with distributed delay.
Suppose the system is in the low metastable state.
Suppose that a burst of protein production reactions fires (Figure~\ref{fig:unfolding}e, top left and top right).
Each of these proteins will enter the population after a production delay.
If the delay is fixed, each burst of reactions will simply be translated in time by the fixed delay, facilitating transitions to the high metastable state (Figure~\ref{fig:unfolding}e, lower left).
However, when the delay is distributed, the bursts break apart, making it harder to transition to the high metastable state (Figure~\ref{fig:unfolding}e, lower right).

We answer the second question by carefully analyzing an extension of the three-states model.

\section{Analysis of the unfolded co-repressive toggle switch when switching rate is low: An extension of the three-states model}
\label{sec:3_states_extension}

\subsection{Review of RM analysis in the fixed-delay case}
\label{subsec:RM_review}

In Section~\ref{subsec:RM_review}, we review the analysis presented in \cite{Gupta-2013-Transcriptional}.
The state space is $\{ H, I, L \}$.
For the co-repressive toggle switch, $H$ and $L$ correspond to neighborhoods of the two metastable states and $I$ corresponds to a thickening of the line of symmetry.
Delay $\tau$ is fixed.
Transition rates for the RM are given by $\lambda_{j \rightarrow k}^{i}$, where $i,j,k \in \{ H, I, L \}$.
Here, $j$ is the current state, $k$ is the target state, and $i$ is the state of the system $\tau$ units of time ago.
The system cannot transition directly from $H$ to $L$ or directly from $L$ to $H$.
The state $I$ must be visited along the way in either case.

We make the following assumptions for the sake of the analysis.
\begin{enumerate}[leftmargin=*, labelindent=\parindent, label=\textbf{(RM\arabic*)}, ref=RM\arabic*]
\item
Fixed delay $\tau$ is small compared to residence times in states $H$ and $L$.
\item
The transition rates out of state $I$ are large compared to the transition rates out of $H$ and $L$.
\item
Define
\begin{equation}
p_{I\rightarrow j}^i=\frac{\lambda_{I\rightarrow j}^i}{\lambda_{I\rightarrow H}^i+\lambda_{I\rightarrow L}^i}.
\end{equation}
We assume
\begin{equation}
p_{I\rightarrow H}^H > p_{I\rightarrow H}^I \text{ and } p_{I\rightarrow L}^L > p_{I\rightarrow L}^I
.
\end{equation}
This assumption reflects the influence of the delay.
\end{enumerate}

Let $R_{H}$ denote residence time in state $H$.
The goal is to estimate $\ev [R_{H}]$.
To do this, we discretize the RM as follows.
The delay $\tau$ is written as $\tau=K\Delta$ where $K\in\N$ and $\Delta= \tau / K$. 
The state $\bm{w} = (w_{-K\Delta}, w_{-(K-1)\Delta}, \ldots , w_{-\Delta},w_0)$ is a $(K+1)$-dimensional vector, where $w_0$ represents the current state of the RM and $w_{-m \Delta}$ represents the state of the RM $m$ steps in the past. 
Restrict jumps to only those that respect the flow of time.
That is, for a jump from $\bm{w}$ to $\bm{u}$, it is necessary that $w_{-(i-1)\Delta}=u_{-i\Delta}$ for all $1 \leqslant i \leqslant K$. 
Note that this severely restricts where $\bm{w}$ can jump.
In particular, a valid jump location $\bm{u}$ is determined entirely by $u_0$. 
The probability of a jump from $\bm{w}$ to $\bm{u}$ is denoted as $\Lambda_{w_0\to u_0}^{w_{-K\Delta}}$. 

Continuing to work with the discretized process, let $f_H$ denote the probability of failing a transition, given that the transition starts in state $H$, and $w_{-i\Delta}=H$ for all $0 \leqslant i\leqslant K$. 
Let $P(k)$ be the probability mass function for the number of steps $k$ that are required to complete the loop $H\to I\to H$, given that $H\to I$ has occurred. 
Then $P(k)$ can be written as 
\begin{equation}
\label{Eq:RM1}
P(k)=\frac1{f_H}\begin{cases}
(\Lambda^{H}_{I\to I})^{k-1}(1-\Lambda^{H}_{I\to I})\frac{\Lambda^{H}_{I\to H}}{\Lambda^{H}_{I\to H}+\Lambda^{H}_{I\to L}} &\text{if } 1 \leqslant k \leqslant K,
\\
(\Lambda^{H}_{I\to I})^K(\Lambda^{I}_{I\to I})^{k-K-1}(1-\Lambda^{I}_{I\to I})\frac{\Lambda^{I}_{I\to H}}{\Lambda^{I}_{I\to H}+\Lambda^{I}_{I\to L}} &\text{if } k>K.
\end{cases}
\end{equation}

So that $P(k)$ is useful for estimating $\E[R_H]$, we assume that before each subsequent transition attempt, the trajectory remains in state $H$ long enough to only remember having been in state $H$. 
Without this assumption, we would have to consider an entire family of conditional probability mass functions, which would make the analysis far less tractable. 
This will come up later when we extend the RM. 

Now we focus on the continuous-time RM.
We continue to let $f_{H}$ be the probability of failing a transition, given that the trajectory remains in state $H$ long enough to only remember having been in state $H$, prior to the occurrence of the $H \rightarrow I$ jump.
Let $F_{H}$ denote the random time needed to complete the $H \rightarrow I \rightarrow H$ loop, given that $H \rightarrow I$ has occurred.
Let $P(t)$ be the probability density function for $F_{H}$.
We find $P(t)$ \revision{by} taking a limit in \eqref{Eq:RM1}.
For fixed $t \in \mathbb{R}_{>0}$ and any $\Delta$ such that $t\Delta^{-1}\in\N$, \eqref{Eq:RM1} gives
\begin{equation}
\label{Eq:RM2}
\begin{aligned}
\Prob &(F_H \in [t-\Delta,t]) =
\\
&\frac1{f_H}\begin{cases}
\big(1-(\lambda_{I\to H}^H+\lambda_{I\to L}^H)\Delta\big)^{t\Delta^{-1}-1}\lambda_{I\to H}^H\Delta &\text{if } \Delta \leqslant t \leqslant \tau ,
\\
\big(1-(\lambda_{I\to H}^H+\lambda_{I\to L}^H)\Delta\big)^{t\Delta^{-1}}\big(1-(\lambda_{I\to H}^I+\lambda_{I\to L}^I)\Delta\big)^{(t-\tau)\Delta^{-1}-1}\lambda_{I\to H}^I\Delta &\text{if } t \geqslant \tau + \Delta .
\end{cases}
\end{aligned}
\end{equation}
Taking $\Delta\to0$, we find that
\begin{equation}
\label{Eq:RM3}
P(t)=\frac1{f_H}\begin{cases}
\lambda_{I\to H}^H\exp\big(-(\lambda_{I\to H}^H+\lambda_{I\to L}^H)t\big) &\text{if } 0<t \leqslant \tau ,
\\
\lambda_{I\to H}^I\exp\big(-(\lambda_{I\to H}^H+\lambda_{I\to L}^H)\tau-(\lambda_{I\to H}^I+\lambda_{I\to L}^I)(t-\tau)\big) &\text{if } t > \tau .
\end{cases}
\end{equation}

We are now in position to compute $f_{H}$.
We have
\begin{subequations}
 \begin{align}
1 &= \int_0^\infty P(t) \, \mathrm{d}t
\\
1 &= \int_0^\infty\frac1{f_H}
\begin{cases}
\lambda_{I\to H}^H\exp\big(-(\lambda_{I\to H}^H+\lambda_{I\to L}^H)t\big) &\text{if } 0 < t \leqslant \tau ,
\\
\lambda_{I\to H}^I\exp\big(-(\lambda_{I\to H}^H+\lambda_{I\to L}^H)\tau-(\lambda_{I\to H}^I+\lambda_{I\to L}^I)(t-\tau)\big) &\text{if } t > \tau
\end{cases}
\, \mathrm{d}t
\\
f_H &= \int_0^\infty
\begin{cases}
\lambda_{I\to H}^H\exp\big(-(\lambda_{I\to H}^H+\lambda_{I\to L}^H)t\big) &\text{if } 0 < t \leqslant \tau ,
\\
\lambda_{I\to H}^I\exp\big(-(\lambda_{I\to H}^H+\lambda_{I\to L}^H)\tau-(\lambda_{I\to H}^I+\lambda_{I\to L}^I)(t-\tau)\big) &\text{if } t > \tau
\end{cases}
\, \mathrm{d}t.
\end{align}
\end{subequations}
Integrating yields
\begin{equation}
    f_H=\Big(1-\exp\big(-(\lambda_{I\to H}^H+\lambda_{I\to L}^H)\tau\big)\Big)\frac{\lambda_{I\to H}^H}{\lambda_{I\to H}^H+\lambda_{I\to L}^H}+\exp\big(-(\lambda_{I\to H}^H+\lambda_{I\to L}^H)\tau\big)\frac{\lambda_{I\to H}^I}{\lambda_{I\to H}^I+\lambda_{I\to L}^I}.
\end{equation}
Using the convenient notation
\begin{equation}
p^i_{I\to H}=\frac{\lambda_{I\to H}^i}{\lambda_{I\to H}^i+\lambda_{I\to L}^i},
\qquad
Z_i(\tau) = \exp\big(-(\lambda_{I\to H}^i+\lambda_{I\to L}^i)\tau\big),
\end{equation}
we arrive at
\begin{equation}
f_H=(1-Z_H(\tau))p^H_{I\to H}+Z_H(\tau)p^I_{I\to H}.
\label{eqn:f_H}
\end{equation}
Therefore $f_H$ is a convex combination of $p^I_{I\to H}$ and $p^H_{I\to H}$, wherein the coefficients are sensitive to $\tau$.

We now complete the estimate for $\ev [R_{H}]$.
Recall that $F_{H}$ is the random time required to complete the $H \rightarrow I \rightarrow H$ loop, given that $H \rightarrow I$ has occurred.
This value is given by
\begin{subequations}
\begin{align}
\E[F_H] &= \int_0^\infty tP(t) \, \mathrm{d}t
\\
&= \left( \frac{1}{f_{H}} \right) \left( \frac{p_{I\to H}^H}{\lambda_{I\to H}^H+\lambda_{I\to L}^H} \right) \Big(1-Z_H(\tau)\big((\lambda_{I\to H}^H+\lambda_{I\to L}^H)\tau+1\big)\Big)
\\
&\qquad {}+ \left( \frac{1}{f_{H}} \right) \left( \frac{p_{I\to H}^I}{\lambda_{I\to H}^I+\lambda_{I\to L}^I} \right) \Big(Z_I(\tau)\big((\lambda_{I\to H}^I+\lambda_{I\to L}^I)\tau+1\big)\Big).
\notag
\end{align}
\end{subequations}
Let $S_{H}$ be the random time required to complete the $H \rightarrow I \rightarrow L$ sequence, given that $H \rightarrow I$ has occurred.
A similar calculation yields $\ev [S_{H}]$.
The derivatives $\mathrm{d} \E[F_H] / \mathrm{d} \tau$ and $\mathrm{d} \E[S_H] / \mathrm{d} \tau$ are very small in magnitude under the assumptions placed on the rates $\lambda^j_{i\to k}$.
Consequently, we may treat $\ev [F_{H}]$ and $\ev [S_{H}]$ as independent of $\tau$ in the final estimate for $\ev [R_{H}]$.

Let $N$ be the number of failed transitions before a successful one (the number of $H \rightarrow I \rightarrow H$ loops that occur before the first $H \rightarrow I \rightarrow L$ sequence occurs).
The random variable $N$ has a geometric distribution with probability mass function
\begin{equation}
\Prob (N=n) = f_{H}^{n} (1 - f_{H})
\end{equation}
for $n \in \mathbb{Z}_{\geqslant 0}$.

The final estimate for $\ev [R_{H}]$ is
\begin{equation}
\label{eqn:Bistab_three-states-approx}
\E[R_H]\sim\frac{f_H}{1-f_H}\left(\E[F_H] + \frac{1}{\lambda_{H \rightarrow I}^{H}} \right)+\E[S_H] + \frac{1}{\lambda_{H \rightarrow I}^{H}}.
\end{equation}

\subsection{Extending the three-states model to explain why the unfolded co-repressive toggle switch destabilizes for low switching rates}
\label{subsec:SH_switching-3-states}

In Figure~\ref{fig:unfolding}a, we see that when switching rate $r$ is low, mean residence time decreases along horizontal slices from left to right.
See Figure~\ref{fig:unfolding}d (blue) for one particular horizontal slice.
To explain this destabilization effect for slow switching rates, we extend the three-states model as follows.
After each transition attempt (jump from $H$ to $I$) has completed, we draw a delay value $\tau$ from the symmetric Bernoulli distribution supported on $\{ \mu - \sigma , \mu + \sigma \}$.
We use this delay value for the RM analysis until the $H \rightarrow I \rightarrow H$ loop or $H \rightarrow I \rightarrow L$ sequence completes.
This allows us to use \eqref{eqn:f_H} one transition attempt at a time.

Let $(M_{i})_{i=1}^{\infty}$ be an independent sequence of random variables, each with the symmetric Bernoulli distribution supported on $\{ 0,1 \}$.
Referring to \eqref{eqn:f_H}, define $p_{0} = f_{H} (\mu - \sigma )$ and $p_{1} = f_{H} (\mu + \sigma )$.
Let $N_{s}$ denote the number of failed transitions before a successful one for the extended RM.

Just as $N$ plays an important role in the analysis of the original RM, $N_{s}$ is important for the extended RM.
$N_{s}$ has conditional probability mass function
\begin{equation}
\Prob (N_{s} = n | (M_{i}) = (m_{i}) ) = (1 - p_{m_{n+1}}) \prod_{1 \leqslant i \leqslant n} p_{m_{i}}.
\end{equation}
A conditioning exercise shows that
\begin{equation}
\label{eqn:Bistab_geometric-limit-commutation}
\E[N_{s}] = \left( \frac{p_0+p_1}{2} \right) \left( 1 - \frac{p_0+p_1}{2} \right)^{-1} .
\end{equation}
Inserting \eqref{eqn:Bistab_geometric-limit-commutation} into \eqref{eqn:Bistab_three-states-approx} gives our estimate for $\ev [R_{H}]$ for the extended RM, namely
\begin{equation}
\E[R_H]\sim\frac{\frac{f_H(\mu-\sigma)+f_H(\mu+\sigma)}2}{1-\frac{f_H(\mu-\sigma)+f_H(\mu+\sigma)}2}\left(\E[F_H]+\frac1{\lambda^H_{I\to H}}\right)+\E[S_H]+\frac1{\lambda^H_{I\to H}}.
\end{equation}
Since $\ev [F_{H}]$ and $\ev [S_{H}]$ are insensitive to delay value, it remains to examine how
\begin{equation}
\label{eq:geometric_sum}
\frac{\frac{f_H(\mu-\sigma)+f_H(\mu+\sigma)}2}{1-\frac{f_H(\mu-\sigma)+f_H(\mu+\sigma)}2}
\end{equation}
behaves as $\sigma$ increases.

First, we note that the function $x\mapsto\frac x{1-x}$ is a monotonically increasing function for $x\in[0,1)$, so the response of $\frac{f_H(\mu-\sigma)+f_H(\mu+\sigma)}{2}$ to changes in $\sigma$ is qualitatively the same as the response of \eqref{eq:geometric_sum}. 
Expanding, we have 
\begin{equation}
\label{eq:expansion_of_fH_average}
\frac{f_H(\mu-\sigma)+f_H(\mu+\sigma)}{2}=p^H_{I\to H}-\left(p^H_{I\to H}-p^I_{I\to H}\right)\left(\frac{Z_H(\mu-\sigma)+Z_H(\mu+\sigma)}{2}\right).
\end{equation}
Noting that $p_{I\to H}^H>p_{I\to H}^I$, the right side of \eqref{eq:expansion_of_fH_average} is a monotonically decreasing function of $\frac{Z_H(\mu-\sigma)+Z_H(\mu+\sigma)}{2}$. 
But $Z_H(\tau)=\exp \left( -(\lambda_{I\to H}^H+\lambda_{I\to L}^H)\tau \right)$ is a strictly convex function, so it follows by Jensen's inequality that
\begin{equation}
\frac{\mathrm{d}}{\mathrm{d} \sigma} \left( \frac{Z_H(\mu-\sigma)+Z_H(\mu+\sigma)}{2} \right) > 0
\end{equation}
for all $\sigma \geqslant 0$.
Backtracking, we have that $\frac{Z_H(\mu-\sigma)+Z_H(\mu+\sigma)}{2}$ strictly increases with $\sigma$, $f_H$ strictly decreases with $Z_H$, and $\frac{f_H}{1-f_H}$ strictly increases with $f_H$, so we conclude that
\begin{equation}
\frac{\mathrm{d}}{\mathrm{d} \sigma} \left( \frac{\frac{f_H(\mu-\sigma)+f_H(\mu+\sigma)}{2}}{1-\frac{f_H(\mu-\sigma)+f_H(\mu+\sigma)}{2}} \right) < 0
\end{equation}
for all $\sigma \geqslant 0$.

Therefore the extended three-states model explains why in Figure~\ref{fig:unfolding}a, we see that when switching rate $r$ is low, mean residence time decreases along horizontal slices from left to right.

\section{Discussion}

In this paper, we have answered a question posed by Kyrychko and Schwartz~\cite{Kyrychko2018} by studying how distributed delay impacts the dynamics of bistable genetic circuits.
We have shown that for a variety of circuits that exhibit bistability, increasing the noise level in the delay distribution dramatically stabilizes the metastable states.
By this we mean that mean residence times in the metastable states dramatically increase.

We have used two methods to explain this stabilization phenomenon.
First, we have introduced and simulated stochastic hybrid models that depend on a switching-rate parameter.
These stochastic hybrid models have allowed us to unfold the distributed-delay models in the sense that, in certain cases, the distributed-delay model can be viewed as a fast-switching limit of the corresponding stochastic hybrid model.
Second, we have generalized the three-states model, a symbolic model of bistability, and analyzed this extension.

Viewed together, the current paper and \cite{Gupta-2013-Transcriptional} show that both delay mean and delay variance can tune the stability of bistable genetic switches.
Holding delay variance fixed and increasing delay mean causes mean residence time to increase because the probability of failing to transition conditioned on attempting to transition goes up.
By contrast, holding delay mean fixed and increasing delay CV causes mean residence time to exhibit a unimodal response (first increase, then decrease) because the additional noise in the delay distribution `smears out' bursts of transcriptional activity, thereby reducing the rate at which transition attempts occur.

Biological switches should ideally switch if and only if it is advantageous to do so.
That is to say, small environmental fluctuations should not cause undesirable switching.
That multiple aspects of the delay distribution -- mean and variance -- can tune the stability of bistable genetic switches is therefore desirable.
This is true from the evolutionary point of view, and with an eye on forward engineering synthetic genetic switches.
We speculate that evolution may have tuned noise levels associated with bistable genetic switches so as to optimize switch stability.
This speculation is reminiscent of ideas from stochastic resonance (noise can act constructively).
The observation that delay mean and delay variance both tune switch stability presents challenges for inference, however.
Delay mean and delay variance would not be identifiable from measurements of residence times alone, for example.

Our work suggests two natural questions with respect to the distributed delay associated with protein production.
First, to what extent can delay mean and delay variance be forward-engineered in synthetic switch designs?
Second, can the delay distribution be derived from detailed models of protein production, such as a totally asymmetric simple exclusion process \cite{Mahima2023}?

There exist several directions for future mathematical research.
First, it would be useful to \revision{develop large deviations theory for bistable genetic switches in the context of distributed delay.
See~\cite{Lv_2014_Constructing, Chaudhury_2015_Moeling, Assaf_2011_Determining} for examples of work on this problem when explicit delay is absent.}
Second, how does distributed delay impact the flow of information through large networks?
Finally, we have used our stochastic hybrid modeling framework in only one specific way in the current paper.
We envision developing a rigorous mathematical foundation for this framework.
It would be interesting to allow the dynamics of the Markov chain to depend on the system that the Markov chain drives.

Overall, our work contributes to the rich literature on the constructive roles of noise in stochastic systems and motivates further study of our stochastic hybrid modeling framework.

\section{Appendix}

\subsection{Further reading}
For a biologist who wants an entry point into how bistable dynamics arise in an experimental setting, and further, an overview of mathematical models that give rise to commonly seen dynamics in biology, we recommend the following two articles:
\begin{enumerate}[leftmargin=*, label=(R\arabic*), ref=R\arabic*, topsep=1ex, itemsep=0.5ex, series=reading]
    \item \textit{Construction of a genetic toggle switch in} Escherichia coli, T.\ S.\ Gardner, C.\ R.\ Cantor, and J.\ J.\ Collins (2000) \cite{gardner2000construction}.\label{R:Gardner}
    \item \textit{Sniffers, buzzers, toggles and blinkers: Dynamics of regulatory and signaling pathways in the cell}, J.\ J.\ Tyson, K.\ C.\ Chen, and B.\ Novak (2003) \cite{tyson2003sniffers}.\label{R:Tyson}
\end{enumerate}
The suggestion \ref{R:Gardner} is a seminal work where Gardner et al.\ construct a synthetic bistable GRN and provide theory that predicts the conditions necessary for bistability. \ref{R:Tyson} is a review of mathematical models that show how simple building blocks in GRNs, such as positive and negative feedback, combine to produce complex dynamics. 

For those who are unfamiliar with the area of stochastic simulation algorithms or large deviations theory, we recommend the following reading:
\begin{enumerate}[resume*=reading]
    \item \textit{Modeling and simulating chemical reactions}, D.\ J.\ Higham (2008) \cite{higham2008modeling}.\label{R:Higham}
    \item \textit{Modeling delay in genetic networks: From delay birth-death processes to delay stochastic differential equations}, C.\ Gupta, J.\ M.\ L\'opez, R.\ Azencott, M.\ R.\ Bennett, K.\ Josi\'c, and W.\ Ott (2014) \cite{gupta2014modeling}.\label{R:Gupta}
    \item \textit{Random perturbations of dynamical systems}, M.\ I.\ Freidlin and A.\ D.\ Wentzell (2012) \cite{freidlin1998random}.\label{R:Freidlin}
    \item \textit{A weak convergence approach to the theory of large deviations}, P.\ Dupuis and R.\ S.\ Ellis (1997) \cite{dupuis1997weak}. \label{R:Dupuis}    
\end{enumerate}
\ref{R:Higham} and \ref{R:Gupta} are, respectively, a SIAM Review of the modeling hierarchy for chemical reaction networks, and an extension of this hierarchy to delay systems that we discuss in this paper. \ref{R:Freidlin} and \ref{R:Dupuis} are rigorous treatments of the large deviations theory of rare events, such as the switching between metastable states observed in simulations of bistable chemical reaction networks. 

The two papers that are most related to ours are:
\begin{enumerate}[resume*=reading]
    \item \textit{Transcriptional delay stabilizes bistable gene networks}, C.\ Gupta, J.\ M.\ L\'opez, W.\ Ott, K.\ Josi\'c, and M.\ R.\ Bennett (2013) \cite{Gupta-2013-Transcriptional}.\label{R:Gupta2}
    \item \textit{Enhancing noise-induced switching times in systems with distributed delays}, Y.\ N.\ Kyrychko and I.\ B.\ Schwartz (2018) \cite{Kyrychko2018}.\label{R:Kyrychko}
\end{enumerate}
\ref{R:Gupta2} shows that increasing mean delay (with 0 variance) increases mean residence times of bistable GRNs. Additionally, \ref{R:Gupta2} introduces the three-states model that we extend in this work. \ref{R:Kyrychko} finds that increasing noise in the delay distribution stabilizes a system with one saddle node and one metastable state, though they incorporate noise in a different way from our paper.

\subsection{Review of delayed SSA}\label{dSSAalg}
We used a modified version of Gillespie's Stochastic Simulation Algorithm (SSA) that incorporates delayed reactions (dSSA) for sampling from the stochastic process presented in \revision{Schlicht} and Winkler, 2008 \cite{Schlicht-2008-delay} to generate the data reported in Figures ~\ref{fig:backstory}-\ref{fig:phage}. 
Here, we present a high level overview of the dSSA. 

Let $J$ be a set indexing the molecular species in the system and let $X(t)\in\Z^J$ be a vector of molecule counts at time $t$.
Further, let $S$ be a finite set representing the types of reactions that can affect $X(t)$. 
As in Gillespie's SSA, reaction $s\in S$ initiates in time interval $[t, t+\Delta t)$ with probability $a_s(x, t)\Delta t$. 
Each reaction has an associated state change vector, $v_s$, that defines the way that the reaction $s$ changes the number of molecules. 
To incorporate delay, we define a delay distribution $d_s(t)$ associated with reaction $s$. 
\revision{Biochemically, reactions that complete after a time delay utilize cellular resources for the duration of the reaction. However, we do not keep track of those resources in the models we consider in this paper.}
Pseudocode for the algorithm is found in Algorithm~\ref{alg:dSSA}. 

Intuitively, the algorithm proceeds as follows: at time 0, initialize a state vector $X$ and define an empty queue. 
Propose the next reaction time $\iota_1$ and select the next reaction $s_1$ to occur, as in Gillespie's SSA. 
Sample a delay $\tau_1$ from $d_s(\iota_1)$, place the reaction completion time and state change vector $(\iota_1+\tau_1, v_{s_1})$ in the queue, and evolve system time to $\iota_1$.
In the second step, propose a new reaction initiation time $\iota_2$. 
If this initiation occurs before the earliest scheduled reaction in the queue, then select the reaction that will occur, sample a delay from the delay distribution, add the appropriate time and change vector to the queue, and evolve the system time to $\iota_2$. 
If this initiation does not occur before the next scheduled reaction, we interpret the event $\iota_2>\iota_1+\tau_1$ as ``no further initiation events occurred in the time interval $[\iota_1,\iota_1+\tau_1)$,'' and discard $\iota_2$. 
Instead of initiating a new reaction, we evolve the system time to $\iota_1+\tau_1$ and increment $X$ by $v_{s_1}$.
Repeat these steps until reaching a desired stopping time.

\subsection{Description of stochastic hybrid modeling framework}\label{sec:stochastic_hybrid}

Here we provide a general description of the modification we made to the standard dSSA to generate the data visualized in Figure~\ref{fig:unfolding}. 
The extension can be stated at a high level as follows: augment the deterministic time dependence of delay distributions with stochastic time dependence. 

Keeping the notation introduced in ~\ref{dSSAalg}, we further introduce $Y(t)$ as a continuous time discrete state Markov chain, and denote by $K$ the finite set of attainable states for the Markov chain. 
We then replace the delay distribution $d_s(t)$ by $d_s(Y,t)$ for each reaction $s\in S$, denoting that this distribution is now dependent on the state of the Markov chain $Y(t)$.   

\begin{figure}[t]
    \centering    \includegraphics[width=.99\textwidth]{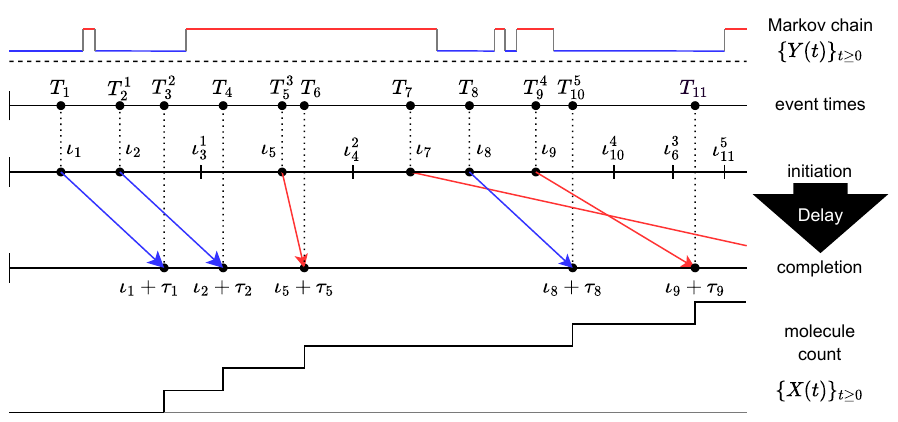}
    \caption{\textbf{Construction of the embedded discrete process.} 
    For illustrative purposes, we show a pure birth process with a two state Markov chain. 
    Each time $\iota_m$ at which a reaction is successfully initiated, a delay value $\tau_m$ is drawn from a distribution corresponding to $Y(\iota_m)$. 
    For example, $\tau_1=\tau_2=\tau_8$ because the delay corresponding to the low (blue) state of the Markov chain is trivially distributed, while the delay distribution for the high (red) state has greater variance. 
    Whenever a reaction initiation time $\iota_m$ is proposed and there is an outstanding reaction scheduled to complete at time $\iota_k+\tau_k<\iota_m$, that initiation at time $\iota_m$ is skipped \revision{in lieu of} completing the reaction at time $\iota_k+\tau_k$. 
        For example, in step $3$, we have $\iota_1+\tau_1<\iota_3$ so $T_3$ becomes $\iota_1+\tau_1$. The same happens at steps $4$, $6$, $10$, and $11$. This figure is adapted from Schlict and Winkler, 2008 \cite{Schlicht-2008-delay}.}
        
    \label{fig:sHybrid}
\end{figure}

The continuous time process $(X(t), Y(t))_{t\geq 0}$ is sampled analogously to the process described in ~\ref{dSSAalg}. 
We will refer to Figure \ref{fig:sHybrid} to visualize the effect of $Y(t)$ on $X(t)$. 
We initialize the process at some state $(x_0, y_0)=(X(0), Y(0))$ with $x_0\in \mathbb{Z}^J$ and $y_0\in K$. 
Denote by $T_1<T_2< \cdots$ the random times at which the Gillespie reactions are either initiated or completed. 
A realization of these times is depicted in the second row of Figure \ref{fig:sHybrid}. 

The values $T_n$ are calculated by the same process as the original dSSA. 
First, a reaction initiation time $\iota_n$ is proposed. There are two possibilities for $\iota_n$ in relation to the reactions already in the queue, enumerated below.
\begin{enumerate}
    \item If $\iota_n$ is less than the next time that a reaction in the queue is scheduled to complete, then the next value in the sequence $(T_n)$ is $\iota_n$. 
    In this case $Y$ is updated independently to time $\iota_n$ if needed.
    Then a delay time, $\tau_n$, is sampled from $d(Y(\iota_n), \iota_n)$ to schedule a completion time, labeled $\iota_n+\tau_n$ in the fourth row of Figure \ref{fig:sHybrid}. 
    In other words, a reaction was initiated (but not completed) at time $T_n=\iota_n$.
    \item Otherwise, if $\iota_n$ is greater than the next scheduled reaction completion time, the next value in the sequence $(T_n)$ represents this completion event. 
    In other words, $T_n=\iota_i+\tau_i$. 
    As in standard dSSA, the reaction $s_i$ then updates the molecule counts to $X(T_n)=X(T_{n-1})+v_{s_i}$, and $\iota_n$ is discarded. 
\end{enumerate} 
Examples of possibility 2 are highlighted with numerical superscripts in Figure \ref{fig:sHybrid}. 
The superscripts match the discarded values of $\iota_n$ to the time in the process at which $\iota_n$ was sampled.

In our extension of the algorithm, the Markov chain $Y(t)$ changes states independently of the Gillespie reactions $X(t)$. 
When a Gillespie reaction is scheduled during possibility 1, a delay is sampled from the distribution $d(Y,t)$, which is determined by the state of $Y(t)$ at the time the reaction is scheduled. 

Algorithm~\ref{alg:Switching_dSSA} outlines the switching extension to the dSSA. 
In short, when a reaction is initiated at a time that occurs before the next completion time in the queue, we evolve the Markov chain up to this initiation time in order to determine the delay distribution from which to sample the delay in the reaction. 

\subsection{Numerical algorithms}
Here we present pseudocode for the two numerical algorithms summarized in this appendix.

\begin{algorithm}[H]
\small
\caption{Delayed Stochastic Simulation Algorithm (dSSA)}\label{alg:dSSA}
\KwIn{
    $X$: initial state vector \\
    $T_{\text{end}}$: simulation end time \\
    $S$: list of reactions, where each $s \in S$ has: \\
    \quad $\bullet$ $v_s$: state change vector \\
    \quad $\bullet$ $a_s(X,t)$: propensity function \\
    \quad $\bullet$ $d_s(t)$: delay distribution \\
    $Q$: queue\revision{*} of scheduled reactions (ordered by completion time), where each $q \in Q$ has: \\
    \quad $\bullet$ $t_q$: time of completion \\
    \quad $\bullet$ $v_q$: state change vector \\
    \revision{*if Q is empty the next reaction is assumed to have $t_q=\infty$ and $v_q=0$.}
}
\KwOut{List of event times $(t,X)$}

$t \leftarrow 0$\;
\While{$t < T_{\text{end}}$}{
    \hfill\tcp{Compute cumulative sum of propensity functions}
    $C \leftarrow [0, a_1(X,t), a_1(X,t)+a_2(X,t), \ldots, \sum_{s\in S} a_s(X,t)]$\;
    $u_1 \sim \text{Uniform}(0,1)$\tcp*{Propose next initiation time}
    $\nu \leftarrow -\frac{\log(u_1)} {C[\text{last}]}$\;
    $\iota \leftarrow t+\nu$\;
    $q \leftarrow Q[\text{next}]$\tcp*{Check if $\nu$ is valid}
    \eIf{$\iota < t_q$}{
        $t \leftarrow \iota$\tcp*{Valid initiation}
        $u_2 \sim \text{Uniform}(0,1)$\tcp*{Select next reaction}
        Choose $s \in S$ such that $C[s-1] \leq u_2 \cdot C[\text{last}] \leq C[s]$\;
        $\tau \sim d_s(t)$\tcp*{Sample delay from distribution}
        $Q.\text{add}(\{t+\tau, v_s\})$\tcp*{Schedule reaction completion}
    }{
        $t \leftarrow t_q$\tcp*{Invalid - complete reaction instead}
        $X \leftarrow X + v_q$\;
        $Q.\text{remove}(q)$\;
    }
}
\Return list of event times $(t,X)$\;
\end{algorithm}

\begin{algorithm}[H]
\small
\caption{Delayed Stochastic Simulation Algorithm with Switching Delays}\label{alg:Switching_dSSA}
\KwIn{
    $X$: initial state vector \\
    $Y$: initial Markov state \\
    $T_{\text{end}}$: simulation end time \\
    $S$: list of reactions, where each $s \in S$ has: \\
    \quad $\bullet$ $v_s$: state change vector \\
    \quad $\bullet$ $a_s(X,t)$: propensity function \\
    \quad $\bullet$ $d_s(Y,t)$: delay distribution \\
    $Q$: queue\revision{*} of scheduled reactions (ordered by completion time), where each $q \in Q$ has: \\
    \quad $\bullet$ $t_q$: time of completion \\
    \quad $\bullet$ $v_q$: state change vector \\
    \revision{*if Q is empty the next reaction is assumed to have $t_q=\infty$ and $v_q=0$.} \\
    $M$: \revision{$K\times K$} rate matrix for the Markov chain
}
\KwOut{List of event times $(t,X,Y)$ and $(t_m,X,Y)$ sorted by time}

$t \leftarrow 0$\;
$t_m \leftarrow 0$ \tcp*{Time for the Markov chain}
$Y_p \leftarrow Y$ \tcp*{Previous state of Markov chain}

\While{$t < T_{\text{end}}$}{
    \hfill\tcp{Compute cumulative sum of propensity functions}
    $C \leftarrow [0, a_1(X,t), a_1(X,t)+a_2(X,t), \ldots, \sum_{s\in S} a_s(X,t)]$\;
    $u_1 \sim \text{Uniform}(0,1)$\tcp*{Propose next initiation time}
    $\nu \leftarrow -\log(u_1) / C[\text{last}]$\;
    $\iota \leftarrow t+\nu$\;
    $q \leftarrow Q[\text{next}]$\tcp*{Check if $\nu$ is valid}
    \eIf{$\iota < t_q$}{
        $t \leftarrow \iota$\tcp*{Valid initiation}
        $u_2 \sim \text{Uniform}(0,1)$\tcp*{Select next reaction}
        Choose $s \in S$ such that $C[s-1] \leq u_2 \cdot C[\text{last}] \leq C[s]$\;
        
        \hfill\tcp{Update Markov chain state}
        \While{$t_m \leq t$}{
            \hfill\tcp{Compute cumulative sum for Markov transitions}
            $CM \leftarrow [0, M[Y,1], M[Y,1]+M[Y,2], \ldots, \revision{\sum_{1=k\neq Y}^{K}M[Y,k]]}$\;
            $u_3 \sim \text{Uniform}(0,1)$\tcp*{Move to next jump time}
            $t_m \leftarrow t_m + \log(u_3)/M[Y,Y]$\;
            $Y_p \leftarrow Y$\tcp*{Record prior state}
            $u_4 \sim \text{Uniform}(0,1)$\tcp*{Select next Markov State}
            Choose $y$ such that $CM[y-1] \leq u_4 \cdot (-M[Y,Y]) \leq CM[y]$\;
            \eIf{y<\revision{Y}}{
                $Y \leftarrow y$\;
            }{
                $Y \leftarrow y+1$\tcp*{Index correction}
            }
        }
        $\tau \sim d_s(Y_p,t)$\tcp*{Sample delay from distribution}
        $Q.\text{add}(\{t+\tau, v_s\})$\tcp*{Schedule reaction completion}
    }{
        $t \leftarrow t_q$\tcp*{Invalid - complete reaction instead}
        $X \leftarrow X + v_q$\;
        $Q.\text{remove}(q)$\;
    }
}
\Return sorted list of event times $(t,X,Y)$ and $(t_m,X,Y)$\;
\end{algorithm}

\section{Declarations}
\subsection{Authorship and Contributorship}
All authors have made substantial intellectual contributions to the study conception, execution, and design of the work. All authors have read and approved the final manuscript.  In addition, the following contributions occurred:  Conceptualization of study, simulations, and analysis: William Ott, Amanda M.\ Alexander; Simulations: Sean Campbell, Courtney C.\ White; Mathematical analysis: All authors; Writing - original draft preparation: All authors; Writing - review and editing: All authors.
\subsection{Conflicts of interest}
The authors declare there are no conflicts of interest.
\subsection{Data and code availability}
Code available upon request.
\subsection{Funding}
A.\ M.\ A.\ was supported by a fellowship from the Gulf Coast Consortia, on the NLM Training
Program in Biomedical Informatics and Data Science T15 LM007093. C.\ C.\ W.\ was supported by a University of Houston Summer Undergraduate Research Fellowship during Summer 2024.


\FloatBarrier

\bibliographystyle{siamplain}
\bibliography{references}

\end{document}